\documentclass[preprint,prd,showpacs,superscriptaddress,preprintnumbers,floatfix]{revtex4}

\def\pslash{\rlap{\hspace{0.02cm}/}{p}}

\usepackage{amsmath}
\usepackage{amssymb}
\usepackage{bm}
\usepackage{dcolumn}
\usepackage{graphicx}
\usepackage{subfigure}
\begin{document}


\date{\today}

\preprint{RIKEN-QHP-14}

\title{%
Tenth-Order QED Contribution to the Lepton Anomalous Magnetic Moment --
Sixth-Order Vertices Containing an Internal Light-by-Light-Scattering Subdiagram
}


\author{Tatsumi Aoyama}
\affiliation{Kobayashi-Maskawa Institute for the Origin of Particles and the Universe (KMI), Nagoya University, Nagoya, 464-8602, Japan}
\affiliation{Nishina Center, RIKEN, Wako, 351-0198, Japan}

\author{Masashi Hayakawa}
\affiliation{Department of Physics, Nagoya University, Nagoya, Japan 464-8602 }
\affiliation{Nishina Center, RIKEN, Wako, 351-0198, Japan}

\author{Toichiro Kinoshita}
\affiliation{Laboratory for Elementary-Particle Physics, Cornell University, Ithaca, New York, 14853, U.S.A }
\affiliation{Nishina Center, RIKEN, Wako, 351-0198, Japan}

\author{Makiko Nio}
\affiliation{Nishina Center, RIKEN, Wako, 351-0198, Japan}

\begin{abstract}
This paper reports the tenth-order QED
contribution to the lepton $g\!-\!2$ from
the gauge-invariant set, called Set III(c), which consists of 
390 Feynman vertex diagrams  containing an internal fourth-order
 light-by-light-scattering subdiagram.
The mass-independent contribution of Set III(c) 
to the electron $g\!-\!2$ ($a_e$) is 4.9210~(103) 
in units of $(\alpha/\pi)^5$.
The mass-dependent contributions to $a_e$ from diagrams containing a muon loop 
is 0.00370~(37) $(\alpha/\pi)^5$.
The tau-lepton loop contribution is negligible at present.
Altogether the contribution of Set III(c) to $a_e$ is 4.9247~ (104)
$(\alpha/\pi)^5$.
We have also evaluated the contribution of the closed electron loop 
to the muon $g\!-\!2$ ($a_\mu$). The result is 7.435~(134) $(\alpha/\pi)^5$.
The contribution of the tau-lepton loop to  $a_\mu$ is 0.1999~(28)
$(\alpha/\pi)^5$.
The total contribution of various 
leptonic loops (electron, muon, and tau-lepton) of Set III(c)  to 
$a_\mu$ is 12.556 (135)~ $(\alpha/\pi)^5$.

\end{abstract}


\maketitle

\section{Introduction}
\label{sec:intro}

The anomalous magnetic moment $a_e \equiv (g\!-\!2)/2$ of 
the electron has played 
the central role in testing the validity of quantum electrodynamics (QED)
as well as the Standard Model.
On the experimental side,
the latest measurement of $a_e$ by the Harvard group has reached the precision
of $0.24 \times 10^{-9}$ 
\cite{Hanneke:2008tm,Hanneke:2010au}:
\begin{eqnarray}
a_e(\text{HV08})= 1~159~652~180.73~ (0.28) \times 10^{-12} ~~~[0.24 \text{ppb}]
~.
\label{a_eHV08}
\end{eqnarray}
The theoretical prediction thus far consists of 
QED corrections of up to the eighth order
\cite{Kinoshita:2005sm,Aoyama:2007dv,Aoyama:2007mn},
direct evaluation of hadronic corrections 
\cite{Davier:2010nc,Hagiwara:2011af,Krause:1996rf,
Melnikov:2003xd,Bijnens:2007pz,Prades:2009tw,Nyffeler:2009tw}, 
and electroweak corrections 
scaled down from their contributions to the muon $g\!-\!2$  
\cite{Czarnecki:1995sz,Knecht:2002hr,Czarnecki:2002nt}. 
To compare the theory with the measurement 
(\ref{a_eHV08}),
we also need  the value of the fine structure constant $\alpha$
determined by a method independent of $g\!-2\!$ .
The best value of such an $\alpha$ available at present is one obtained 
from the measurement of $h/m_{\text{Rb}}$, the ratio of the Planck constant
and the mass of Rb atom,  
combined with the very precisely known values of the Rydberg constant 
and $m_\text{Rb}/m_e$: \cite{Bouchendira:2010es}
\begin{eqnarray}
\alpha^{-1} (\text{Rb10}) = 137.035~999~037~(91)~~~[0.66 \text{ppb}].
\label{alinvRb10}
\end{eqnarray}  
With this  $\alpha$ the theoretical prediction of $a_e$ becomes 
\begin{eqnarray}
a_e(\text{theory}) = 1~159~652~181.13~(0.11)(0.37)(0.02)(0.77) \times 10^{-12},
\label{a_etheory}
\end{eqnarray}
where the first, second, third, and fourth uncertainties come
from the calculated eighth-order QED term \cite{Aoyama:2007mn}, 
the crude tenth-order estimate \cite{Mohr:2008fa}, 
the hadronic and electroweak contributions, and the
fine structure constant (\ref{alinvRb10}), respectively.
The theory (\ref{a_etheory})
is in good agreement with the
experiment (\ref{a_eHV08}):  
\begin{eqnarray}
a_e(\text{HV08}) - a_e(\text{theory}) = -0.40~ (0.88) \times 10^{-12},
\end{eqnarray}
proving that QED (Standard Model) is in good shape even at this very high
precision.

Eq. (\ref{a_etheory}) shows clearly that the largest source of uncertainty is
the fine structure constant (\ref{alinvRb10}).
To put it differently, a non-QED $\alpha$, even the best one available
at present, is too crude to test QED to the extent achieved by the theory and measurement of $a_e$.
Thus it makes more sense to test QED by an alternate approach,
namely, obtain $\alpha$ from theory and measurement of 
$a_e$\cite{Hanneke:2008tm}:
\begin{eqnarray}
\alpha^{-1}(a_e 08) = 137.035~999~085~(12)(37)(2)(33)~~~[0.37 \text{ppb}],
\label{alinvae}
\end{eqnarray}
where the first, second, third, and fourth uncertainties come
from the calculated eighth-order QED term, the crude tenth-order estimate, 
the hadronic and electroweak contributions, and the
measurement of $a_e$(HV08), respectively.

Although the uncertainty of
$\alpha^{-1} (a_e 08)$ in (\ref{alinvae}) is a factor 2 smaller than
$\alpha^{-1}$(Rb10), it is not a firm factor since it
 depends on the estimate of the tenth-order term,
which is only a crude guess \cite{Mohr:2008fa}.
For a more stringent test of QED, it is obviously necessary to calculate
the actual value of the tenth-order term.
To meet this challenge we launched 
several years ago
a systematic program 
to evaluate the complete tenth-order term
\cite{Kinoshita:2004wi,Aoyama:2005kf,Aoyama:2007bs}.

The 10th-order QED contribution to the 
anomalous magnetic moment of an electron can be written as
\begin{equation}
	a_e^{(10)} 
	= \left ( \frac{\alpha}{\pi} \right )^5 
          \left [ A_1^{(10)}
	+ A_2^{(10)} (m_e/m_\mu) 
	+ A_2^{(10)} (m_e/m_\tau) 
	+ A_3^{(10)} (m_e/m_\mu, m_e/m_\tau) \right ],
\label{eq:ae10th}
\end{equation}
where $m_e/m_\mu = 4.836~331~66~(12) \times 10^{-3}$ and
$m_e/m_\tau = 2.875~64~(47) \times 10^{-4}$ \cite{Mohr:2008fa}.
In the rest of this article the factor 
$ \left ( \frac{\alpha}{\pi} \right )^5$ is suppressed for simplicity.
 
The diagrams contributing to the mass-independent term $A_1^{(10)}$ can be
classified into six gauge-invariant sets, further divided into
32 gauge-invariant subsets depending on the nature of closed
lepton loop subdiagrams.
Thus far, numerical results of 30 gauge-invariant subsets, 
which consist of 5928 vertex diagrams, 
have been published \cite{Kinoshita:2005sm,Aoyama:2008gy,Aoyama:2008hz,Aoyama:2010yt,Aoyama:2010pk,Aoyama:2010zp,Aoyama:2011rm,Aoyama:2011zy},
or submitted for publication \cite{Aoyama:2011dy}.
Five of these 30 subsets are also known analytically \cite{Laporta:1994md,Aguilar:2008qj}.
They are in good agreement with our calculations.

%
\begin{figure}[th]
\resizebox{12cm}{!}{\includegraphics{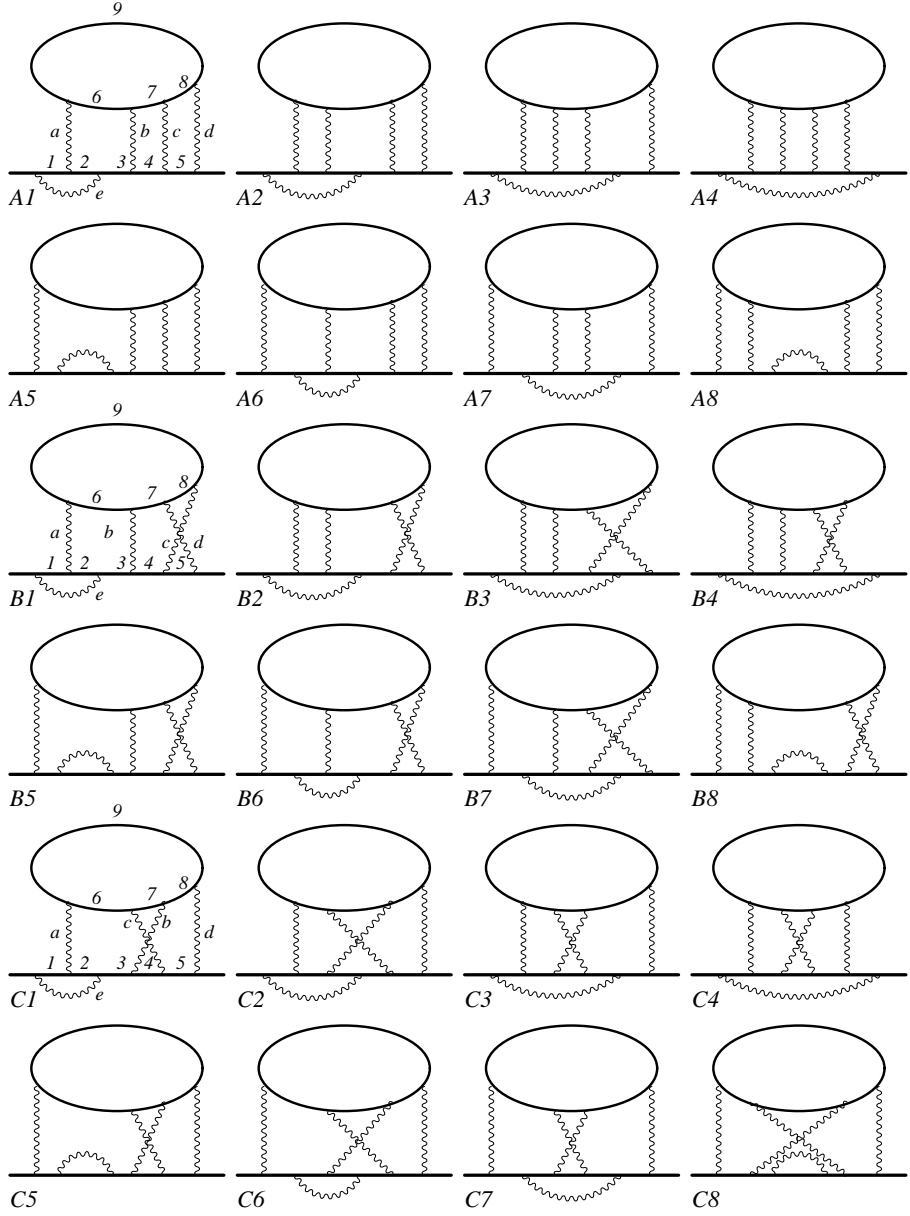}}
\caption{\label{fig:set3c} 
Tenth-order self-energy-like diagrams in which lepton lines propagate
in the magnetic field. They represent 390 vertex
diagrams of Set III(c).
Assignment of Feynman parameters 
$z_1, z_2, \ldots, z_9$ to the lepton lines and
$z_a, z_b, .., z_e$ to the photon lines 
is indicated in the figure $A1$, $B1$, and $C1$.
}
\end{figure}
\begin{figure}[th]
\resizebox{12.0cm}{!}{\includegraphics{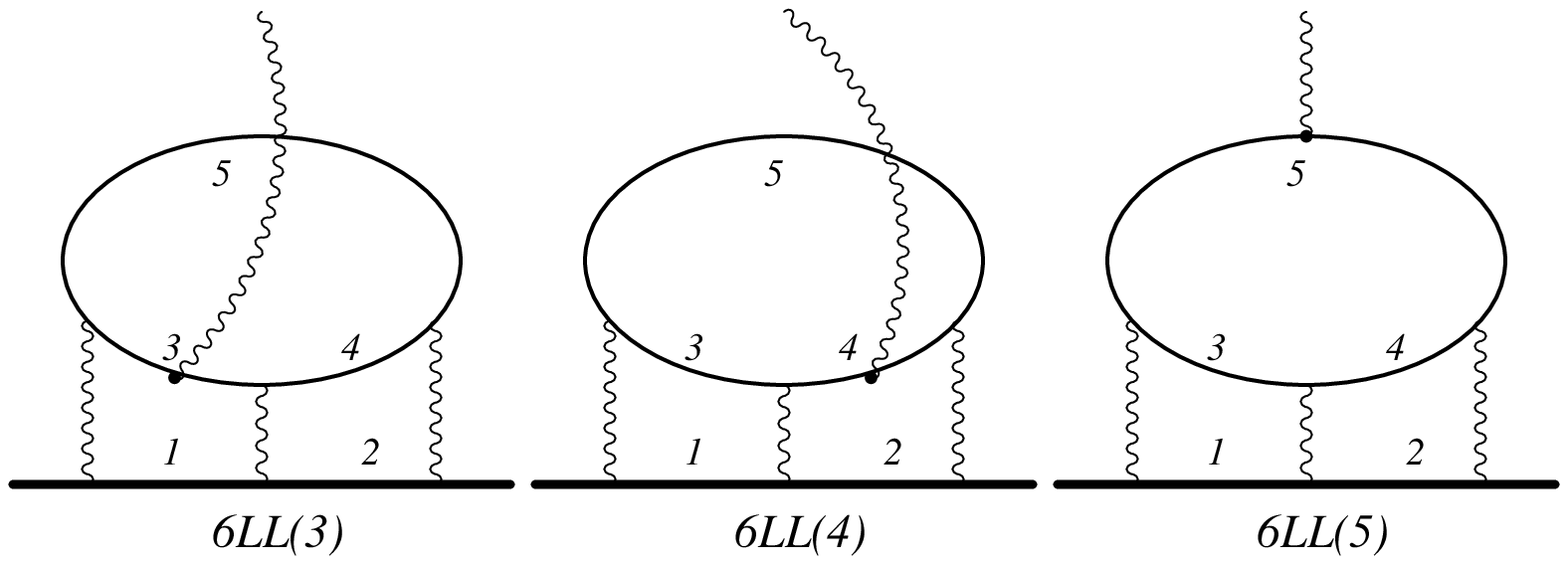}}
\caption{
\label{fig:6LL} 
Vertex diagrams of sixth order containing an {\em l-l} subdiagram.
The diagram $6LL(3)$ and $6LL(4)$ are identical with each other because of the
time-reversal symmetry.
There are six vertex diagrams of this type, taking into account of two directions
in which the closed fermion loop can take.
}
\end{figure}
%

\begin{figure}[t]
\resizebox{16.0cm}{!}{\includegraphics{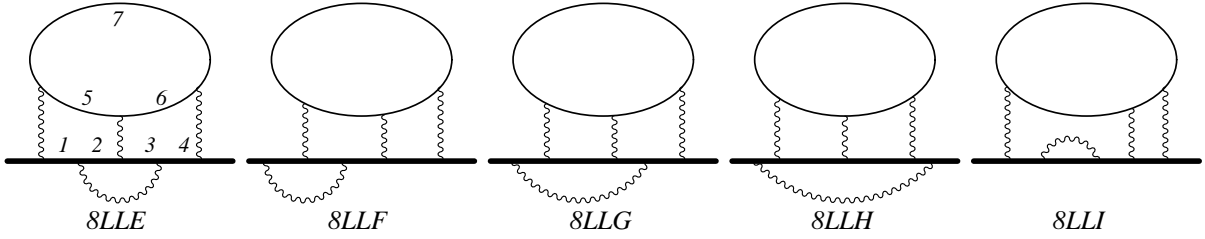}}
\caption{
\label{fig:G4c} 
Various diagrams of the eighth order needed for renormalization.
$8LL\alpha~(\alpha=E,F,G,H,I)$  is denoted as $LL\alpha$  in
Ref.~\cite{Kinoshita:1981ww}. 
A vertex diagram $8LL\alpha(i)~(i=5,6,7)$ 
is obtained by inserting an external vertex into
a fermion line $i$ of the diagram $8LL\alpha$.
It is denoted as $LL\alpha (i)$  in Ref.~\cite{Kinoshita:2002ns}. 
}
\end{figure}

\begin{figure}[t]
\resizebox{12.0cm}{!}{\includegraphics{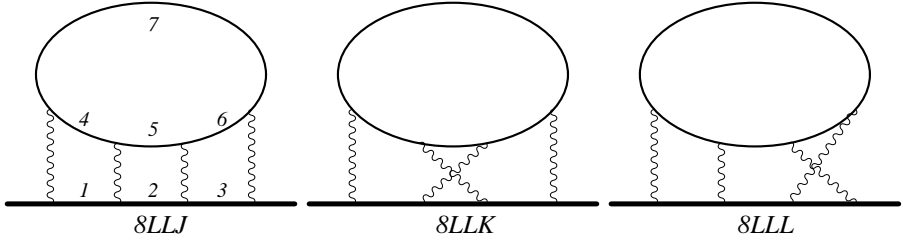}}
\caption{
\label{fig:LLjkl} 
The eighth-order self-energy-like diagrams $8LLJ$, $8LLK$, and $8LLL$, containing a light-by-light
scattering loop internally.
Open lepton lines propagate in the weak magnetic field.
They represent 18 vertex diagrams in total.
The vertex diagram $8LLJ(1)$ is a part of the diagram $8LLJ$
in which the magnetic vertex is attached only to the fermion line 1 of
the self-energy diagram $8LLJ$.
}
\end{figure}

In this paper we report the contribution to $A_1^{(10)}$  from 
the gauge-invariant subset called Set III(c), which consists of
390 vertex diagrams and is represented by 24 self-energy-like
diagrams of Figure~ \ref{fig:set3c}.
A characteristic feature of these diagrams 
is that they have vertex and self-energy subdiagrams which contain
a light-by-light ({\em l-l}) scattering subdiagram.
They can be classified into three types:
%
\begin{enumerate}
  \item  Sixth-order vertex subdiagrams containing an {\em l-l} 
loop {\it externally} as is shown in Figure~\ref{fig:6LL}. 
Here ``external" means that one of the photons is external to the 
subdiagrams containing the {\em l-l} loop.
  \item Eighth-order vertex subdiagrams containing an {\em l-l} 
loop externally as is shown in Figure~\ref{fig:G4c}. They are obtained
by applying a virtual photon correction on the open fermion line of a 
sixth-order {\em l-l} vertex subdiagram of Figure~\ref{fig:6LL}. 
  \item  Eighth-order vertex and self-energy subdiagrams 
which contain an {\em l-l} loop internally. 
This type appears for the first time in the tenth-order 
perturbation theory of QED.  See Figure~\ref{fig:LLjkl}. 
\end{enumerate}
%

The vertex subdiagrams of type 1 and type 2 do not have 
their Ward-Takahashi-related self-energy subdiagrams  which 
vanish identically due to Furry's theorem.
Thus, the gauge-invariant sums of the vertex renormalization constants 
of these external {\em l-l} subdiagrams  also vanish identically.

For vertex subdiagrams of type 3, the corresponding self-energy subdiagrams do exist 
which have an internal {\em l-l} diagram. 
In this case both vertex subdiagram and self-energy subdiagram  
have UV divergence due to 
the sixth-order external {\em l-l} vertex subdiagram. 

Because of these specific features of an {\em l-l} scattering diagram 
and vertex diagrams containing an external {\em l-l} loop, 
we adopt for the Set III(c) an approach 
different from the one used for a diagram without an {\em l-l} loop
\cite{Aoyama:2005kf,  Aoyama:2007bs}.
Our formulation and treatment of UV divergences and IR divergences
due to subdiagrams are described in Sec.~\ref{sec:formulation}.
Results of numerical evaluation will be presented in Sec.~\ref{sec:numerical}.
Sec.~\ref{sec:discussion} is devoted to the summary and discussion 
of this work.
Renormalization of these diagrams is described in Appendix
\ref{sec:renorm}.

%

\section{Formulation}
\label{sec:formulation}

Instead of dealing with the 390 vertex diagrams of Set III(c) individually,
we consider the sum $\Lambda^\nu$ of a set of vertex diagrams
that are obtained from a self-energy-like diagram $\Sigma (p)$ 
of Figure~\ref{fig:set3c} 
by inserting a magnetic vertex $\gamma^\nu$ 
in the lepton lines 1, 2, 3, 4, and 5.
We rewrite this $\Lambda^\nu$ as
\begin{equation}
\Lambda^\nu (p, q) \simeq - q^\mu \left [ \frac{\partial \Lambda_\mu (p, q)}
{\partial q_\nu} \right ]_{q=0} - \frac{\partial \Sigma (p)}{\partial p_\nu}.
\label{WTid}
\end{equation}
with the help of the Ward-Takahashi identity,
where $p-q/2$ and $p+q/2$ are the 4-momenta of incoming and outgoing
lepton lines and $(p-q/2)^2 =(p+q/2)^2 = m^2$.
Each sum corresponds to one of the 24 self-energy-like diagrams
shown in Figure~\ref{fig:set3c}.
The $g\!-\!2$ term is projected out from the right-hand side of (\ref{WTid}).

\subsection{Construction of Unrenormalized Integrals}
\label{construction}

Each diagram ${\cal G}$ of Figure~\ref{fig:set3c} can be expressed by
a momentum integral applying the Feynman-Dyson rule.
Introducing Feynman parameters $z_1, z_2, \ldots, z_9$
for the electron propagators and $z_a, z_b, \ldots, z_e$
for the photon propagators (see the figures $A_1$, $B_1$, and $C_1$
 of Figure~\ref{fig:set3c}),
we carry out the momentum integration analytically
by means of a home-made program written in FORM
\cite{Vermaseren:2000nd}.
This leads to an integral of the form
\begin{equation}
M_{\cal G} = - \left (\frac{-1}{4} \right )^5 4! \int (dz)_{\cal G}
 \left[\frac{1}{4} \left ( \frac{E_0 + C_0}{U^2V^4} 
                        +  \frac{E_1+C_1}{U^3V^3} + \cdots \right) 
                 + \left (\frac{N_0+Z_0}{U^2 V^5} 
                       +  \frac{N_1+Z_1}{U^3 V^4}+\cdots \right) \right],
\label{M10}
\end{equation}
where ${E_n}, {C_n}, {N_n}$ and ${Z_n}$ 
are functions of Feynman parameters,
and ``symbolic" building blocks $A_i, B_{ij}, C_{ij}$,
for $i, j = 1, 2, \ldots, 9$.
$n$ is the number of contractions (see \cite{Kinoshita:1990} for definitions).
$U$ is the Jacobian of transformation from the momentum space variables to
Feynman parameters. 
$A_i$ is the {\it scalar current} defined by
\begin{equation}
A_i = \eta_{i} - \frac{1}{U} \sum_{j=1}^5 z_j B_{ij}, 
~~~~~
\left\{ 
\begin{array}{lll}
\eta_i = 1 & \text{for} & i=1,2,3,4,5 \\
\eta_i = 0 & \text{for} & i=6,7,8,9 
\end{array}
\right . 
,
\end{equation}
and 
\begin{equation}
(dz)_{\cal G} = \prod_{i \in {\cal G}} dz_i (1- \sum_{i \in {\cal G}} z_i ).
\end{equation}
See, for example, Ref. \cite{Aoyama:2005kf} for definitions of $B_{ij}$ and $C_{ij}$.
$V$ is obtained by combining all denominators of propagators into one
with the help of Feynman parameters.
It has a form common to all diagrams of Figure~\ref{fig:set3c}:
\begin{equation}
V= \sum_{i=1}^9 z_i (1 - A_i) m_i^2 + \sum_{k=a}^e z_k \lambda_k^2,
\label{eq:defV}
\end{equation}
where $m_i$ and $\lambda_k$ are the rest masses of lepton $i$
and photon $k$, respectively.
Of course, $m_i$ is independent of $i$ and $\lambda_k$ is 0 independent of $k$.
But it is useful to distinguish different lepton lines and
photon lines in deriving Eq. (\ref{M10}).
The form of $A_i$ as a function of Feynman parameters depends
on the structure of individual diagram ${\cal G}$ of Figure~\ref{fig:set3c}.

\subsection{Renormalization}
\label{renormalization}

The diagrams of Set III(c) as a whole form a (formal) gauge-invariant set.
However, individual diagrams have UV divergences arising
from the light-by-light-scattering ({\em l-l}) subdiagram
as well as vertex subdiagrams or self-energy subdiagrams.
All these divergences must be regularized in advance.
In order to maintain gauge invariance the {\em l-l} subdiagram
may be regularized by the Pauli-Villars method 
or by the dimensional regularization.
The sixth-order vertex renormalization constant associated with the diagram
containing an {\em l-l} subdiagram and the eighth-order vertex renormalization constant
containing an {\em l-l} subdiagram are logarithmically divergent,
but their sum over all diagrams vanishes due to Ward-Takahashi identity.  
Note that the self-energy diagrams associated with these vertex diagrams
do not exist in QED because of the Furry's theorem. 

As is indicated in the figures $A1$, $B1$, $C1$ of Figure~\ref{fig:set3c},
we denote open fermion lines as 1, 2, 3, 4, 5,
fermion lines forming a closed loop as 6, 7, 8, 9, and photon lines as a, b, c, d, e. 
We will 
identify a subdiagram containing open lepton lines in terms of 
a subset of (1,2,3,4,5).
For instance, the vertex subdiagram (1,2) of $A1$ will be 
denoted by  (1,2), and
the vertex subdiagram \{2,3,4,5; 6,7,8,9; b,c,d\} of $A1$ will be denoted
by  (2,3,4,5).
An exception is the {\em l-l} subdiagram, which will be denoted as (6,7,8,9).
Under this convention the diagram $A1$ has five divergent subdiagrams
$(1,2)$, $(4,5)$, $(1,2,3,4)$, $(2,3,4,5)$, and $(6,7,8,9)$. 
The fifteen UV subtraction terms can be constructed from these subdiagrams
following the Zimmermann's forest formula \cite{Zimmermann:1969}. 


Diagrammatically, the second-order vertex subdiagram appears not only in the forests including the subdiagram $(1,2)$
but also in the forest $(2,3,4,5)(4,5)$. In the latter, the reduced diagram $(2,3)$ forms a second-order vertex diagram.  We will treat renormalization 
of this {\it implicit} second-order vertex in a manner different 
from the {\it explicit} second-order vertex. A detailed account will be given in Appendix A.

The UV divergence arising from the explicit second-order vertex (1,2) of 
the diagram $A1$ can be subtracted
by an integral defined by the ${\mathit K}_{12}$-operation
\cite{Kinoshita:1990} applied on the integral $M_{A1}$.
The ${\mathit K}_{12}$-operation is defined in such a way that the
result of the operation factorizes exactly as
\begin{equation}
{\mathit K}_{12} M_{A1} = L_2^{(1,2)UV} M_{8LLJ}^{(3,4,5)},
\end{equation}
where $L_2^{(1,2)UV}$ is the UV-divergent part of the second-order 
on-shell vertex renormalization constant $L_2^{(1,2)}$
and $M_{8LLJ}^{(3,4,5)}$ is the magnetic moment amplitude from the eighth-order self-energy-like 
diagram $8LLJ$ of Figure~\ref{fig:LLjkl}.

UV divergences from the explicit second-order vertex subdiagram
are also found in the diagrams $B1, C1, A6, B6,$ and  $C6$.
UV divergences due to the explicit second-order self-energy-like 
subdiagram come from the diagrams $A5, B5, C5, A8, B8,$ and  $C8$.
The renormalization scheme in which only these second-order divergences
appear are handled by the ${\mathit K}$-operation
and is described in Appendix \ref{sec:renorm}.

All other subdiagrams contain an {\em l-l} subdiagram,
which we treat by the Pauli-Villars method or
by the dimensional regularization.
For instance, in the latter method, let $F_{\alpha i} (d)$ be one of such integrals 
defined in $d$ dimension,
where ${\alpha i}$ takes values $\alpha = A, B, C;~ i=1, 2, ...,8$.
Let $G_{\alpha i} (d)$ be $F_{\alpha i} (d)$ in which the {\em l-l} subdiagram 
(of the form $\Pi_{\mu \nu \sigma \rho} (k_1, k_2, k_3, k_4)$)
is replaced by
the tensor with zero external momenta, namely,
$\Pi_{\mu \nu \sigma \rho} (0, 0, 0, 0)$.
Let us rewrite $F_{\alpha i} (d)$ symbolically as
\begin{equation}
[F_{\alpha i} (d) - G_{\alpha i} (d)] + G_{\alpha i} (d),
\end{equation}
where by ``symbolically" we mean that subtraction is performed on the integrand
before the integration is carried out.
Now we can safely take the limit $d \rightarrow 4$ for the term
$[F_{\alpha i} (d) - G_{\alpha i} (d)]$
since its integrand does not cause UV divergence.
Of course, the second term $G_{\alpha i} (d)$ is singular for $d \rightarrow 4$.
However, gauge invariance guarantees that the sum of $G_{\alpha i} (d)$
over all diagrams of Figure~\ref{fig:set3c} vanishes for any value of dimension $d$:
\begin{equation}
 \sum_{\alpha = A}^C \sum_{i=1}^8 \eta_i G_{\alpha i} (d) = 0,
\end{equation}
where $\eta_i = 2 $ for $i=4, 7, 8$, 
and $\eta_i = 4 $ for $i=1, 2, 3, 5, 6$.
Thus, in the end, we have to compute only 
\begin{equation}
\lim_{d \rightarrow 4} [F_{\alpha i} (d) - G_{\alpha i} (d)].
\label{FminusG}
\end{equation}
Of course the same result is obtained by the Pauli-Villars method.
To avoid crowded notations let us use $F_{\alpha i} (4)$ instead of Eq. (\ref{FminusG})
in the following.

Each self-energy-like diagram of Figure~\ref{fig:set3c} represents 
the sum of five vertex diagrams.
Diagrams obtained by reversing the momentum flow within 
the {\em l-l} loop are not shown
but they give the same integrals as the original ones.
Another factor 2 must be included for diagrams that are not symmetric
under time-reversal.
Thus, integrals for diagrams such as $A1$ actually represent 
$2 \times 2 \times 5$ vertex diagrams.
The $g\!-\!2$ contribution from 
the sum of all diagrams of Set III(c),
after the renormalization described in Appendix \ref{sec:renorm}
is carried out, can thus be written as
\begin{equation}
A_1^{(10)} [\text{Set~III(c)}^{(l_1l_2)}] 
 = \sum_{\alpha = A}^C \sum_{i=1}^8 \eta_i \Delta M_{\alpha i}^{(l_1l_2)}
        -3 \Delta L\!B_2 \Delta M_{8JKL}^{(l_1l_2)} ,
\label{eq:set3c}
\end{equation}
where $l_1$ refers to the open lepton line and $l_2$ refers to the closed lepton
line.
$\eta_i = 2$ for $i = 4, 7, 8$,  and $\eta_i = 4$ for $i= 1, 2, 3, 5, 6$.
$\Delta LB_2$ and $\Delta M_{8JKL}^{(l_1l_2)}$ are defined in 
Appendix \ref{sec:renorm}.
%
 

\section{Numerical results}
\label{sec:numerical}

Evaluation of integral $\Delta M_{\alpha i}^{(l_1 l_2)}$ is carried out by 
the adaptive-iterative Monte-Carlo
 integration routine VEGAS \cite{Lepage:1977sw}.
The results for the case $(l_1 l_2)=(ee)$ are listed in Table \ref{table:setIII(c)_(ee)}.
From this table and Table \ref{table:renom} listing the residual renormalization terms we obtain
\begin{equation}
A_1^{(10)} [\text{Set~III(c)}^{(ee)}] = 4.9210~(103).
\label{set3c_ee}
\end{equation}

The contribution of the muon loop to $a_e$ can be calculated from the data
listed in 
Table \ref{table:setIII(c)_(em)} 
and Table \ref{table:renom}:
\begin{equation}
A_2^{(10)} [\text{Set~III(c)}^{(em)}] = 0.00370~(37).
\label{set3c_em}
\end{equation}

The contribution of the tau-lepton loop to $a_e$ 
is within the uncertainty of (\ref{set3c_ee}).
Thus the total QED contribution to $a_e^{(10)}$
is essentially the sum of (\ref{set3c_ee}) and (\ref{set3c_em}):
\begin{equation}
a_e^{(10)} [\text{Set~III(c)}] = 4.9247~(104)
            \left ( \frac{\alpha}{\pi} \right)^5.
\label{set3c_eall}
\end{equation}
%

%

FORTRAN programs for $a_e$ can be readily adapted to the evaluation of $a_\mu$.
The results of evaluation
of the contribution of the electron loop to the muon $g\!-\!2$
are listed in Table \ref{table:setIII(c)_(me)}.  From this table and 
Table \ref{table:renom} we obtain
\begin{equation}
A_2^{(10)} [\text{Set~III(c)}^{(me)}] = 7.435~(134).
\label{set3c_me}
\end{equation}
The contribution of the tau-lepton loop to $a_\mu$ 
is calculated from the data
listed in Table \ref{table:setIII(c)_(mt)} and 
Table \ref{table:renom}:
\begin{equation}
A_2^{(10)} [\text{Set~III(c)}^{(mt)}] = 0.1999~(28).
\label{set3c_mt}
\end{equation}
The total QED contribution to $a_\mu^{(10)}$
is the sum of (\ref{set3c_ee}),
(\ref{set3c_me}), and (\ref{set3c_mt}):
\begin{equation}
a_\mu^{(10)} [\text{Set~III(c)}] = 12.556~(135) 
            \left ( \frac{\alpha}{\pi} \right)^5.
\label{set3c_muall}
\end{equation}

\begingroup
\renewcommand{\baselinestretch}{1.0}

\begin{table*}
\caption{%
Contributions of diagrams of Set~III(c) to $a_e$
for $(l_1l_2) = (ee)$.
The superscript $(ee)$ is suppressed for simplicity.
The multiplicity $n_F$ is the number of vertex diagrams
represented by the integral and
is incorporated in the numerical value.
All integrals are evaluated initially with $10^8$ sampling points per iteration, iterated 50 times, followed by $10^9$ points, iterated several times.
  \label{table:setIII(c)_(ee)}
}

\begin{ruledtabular}
\begin{tabular}{lcdcc}
\multicolumn{1}{c}{Integral} &
\multicolumn{1}{c}{$n_F$} &
\multicolumn{1}{c}{\hspace*{4em}\parbox[t]{10em}{Value (Error) \\[-1ex] including $n_F$}} &
\multicolumn{1}{c}{\parbox[t]{10em}{Sampling per \\[-1ex]  iteration}} &
\multicolumn{1}{c}{\parbox[t]{4em}{No. of \\[-1ex] iterations}} \\[3ex]
\hline

$\Delta M_{A1}$ & 20 & -4.255~92~(253) & $ 1\times 10^8,~1\times 10^9 $ & 50,~400\\
$\Delta M_{A2}$ & 20 &  4.938~78~(244) & $ 1\times 10^8,~1\times 10^9 $ & 50,~300\\
$\Delta M_{A3}$ & 20 & -1.546~88~(246) & $ 1\times 10^8,~1\times 10^9 $ & 50,~345\\
$\Delta M_{A4}$ & 10 & -0.323~88~(127) & $ 1\times 10^8,~1\times 10^9 $ & 50,~ 30\\
$\Delta M_{A5}$ & 20 &  6.320~29~(153) & $ 1\times 10^8,~1\times 10^9 $ & 50,~ 60\\
$\Delta M_{A6}$ & 20 & -5.660~33~(218) & $ 1\times 10^8,~1\times 10^9 $ & 50,~300\\
$\Delta M_{A7}$ & 10 &  2.284~61~(173) & $ 1\times 10^8,~1\times 10^9 $ & 50,~ 65\\
$\Delta M_{A8}$ & 10 &  1.362~06~(129) & $ 1\times 10^8,~1\times 10^9 $ & 50,~ 20\\
$\Delta M_{B1}$ & 20 &  5.693~53~(293) & $ 1\times 10^8,~1\times 10^9 $ & 50,~412\\
$\Delta M_{B2}$ & 20 & -7.018~17~(273) & $ 1\times 10^8,~1\times 10^9 $ & 50,~302\\
$\Delta M_{B3}$ & 20 &  3.735~46~(260) & $ 1\times 10^8,~1\times 10^9 $ & 50,~342\\
$\Delta M_{B4}$ & 10 & -0.052~76~(122) & $ 1\times 10^8,~1\times 10^9 $ & 50,~ 30\\
$\Delta M_{B5}$ & 20 & -4.739~40~(166) & $ 1\times 10^8,~1\times 10^9 $ & 50,~ 60\\
$\Delta M_{B6}$ & 20 &  3.061~01~(212) & $ 1\times 10^8,~1\times 10^9 $ & 50,~300\\
$\Delta M_{B7}$ & 10 &  0.351~39~(168) & $ 1\times 10^8,~1\times 10^9 $ & 50,~ 65\\
$\Delta M_{B8}$ & 10 & -0.793~52~(136) & $ 1\times 10^8,~1\times 10^9 $ & 50,~ 20\\
$\Delta M_{C1}$ & 20 &  0.377~40~(279) & $ 1\times 10^8,~1\times 10^9 $ & 50,~417\\
$\Delta M_{C2}$ & 20 &  3.054~41~(241) & $ 1\times 10^8,~1\times 10^9 $ & 50,~300\\
$\Delta M_{C3}$ & 20 & -1.329~04~(260) & $ 1\times 10^8,~1\times 10^9 $ & 50,~338\\
$\Delta M_{C4}$ & 10 &  0.435~88~(131) & $ 1\times 10^8,~1\times 10^9 $ & 50,~ 30\\
$\Delta M_{C5}$ & 20 & -3.729~22~(159) & $ 1\times 10^8,~1\times 10^9 $ & 50,~ 60\\
$\Delta M_{C6}$ & 20 &  4.273~41~(258) & $ 1\times 10^8,~1\times 10^9 $ & 50,~300\\
$\Delta M_{C7}$ & 10 & -2.233~00~(159) & $ 1\times 10^8,~1\times 10^9 $ & 50,~ 65\\
$\Delta M_{C8}$ & 10 & -1.514~28~(142) & $ 1\times 10^8,~1\times 10^9 $ & 50,~ 20\\

\end{tabular}
\end{ruledtabular}
\end{table*}

\endgroup

\begingroup
\renewcommand{\baselinestretch}{1.2}

\begin{table}
  \caption{%
Auxiliary integrals for  Set~III(\textit{c}). 
Some integrals are known exactly. 
Other integrals are obtained by the integration routine VEGAS.
The superscript $(l_1~l_2)$ indicates that the open and closed
fermion lines consist of fermions $l_1$ and $l_2$, respectively.
The letters $e$, $m$, and $t$ stand for electron, muon, and tau-lepton,
respectively.
\label{table:renom}
}
\begin{ruledtabular}
\begin{tabular}{ldld}
\multicolumn{1}{c}{Integral} & 
\multicolumn{1}{c}{Value (error)} &
\multicolumn{1}{c}{Integral} & 
\multicolumn{1}{c}{Value (error)} \\[1ex]
\hline
$ M_2$                        &  0.5            &
$\Delta L\!B_2$               &  0.75           \\ 
  & & & \\
$\Delta M_{8JKL}^{(ee)}$      & -0.990~72~( 11)  & 
$\Delta M_{8JKL}^{(me)}$      & -4.432~43~( 59)  \\ 
    &&&   \\ 
$\Delta M_{8JKL}^{(em)}$      & -0.000~177~8~( 13)  & 
$\Delta M_{8JKL}^{(mt)}$      & -0.015~87~( 5)  \\ 
    \end{tabular}
  \end{ruledtabular}
\end{table}

\endgroup

\begingroup
\renewcommand{\baselinestretch}{1.0}

\begin{table*}
\caption{%
Contributions of diagrams of Set~III(c) to $a_e$
for $(l_1l_2) = (em)$.
The superscript $(em)$ is suppressed for simplicity.
The multiplicity $n_F$ is the number of vertex diagrams
represented by the integral and
is incorporated in the numerical value.
All integrals are evaluated  with $10^7$ sampling points per iteration, iterated 50 times,  and subsequently evaluated with $10^8$ sampling points per
iteration, iterated 50 times.
  \label{table:setIII(c)_(em)}
}

\begin{ruledtabular}
\begin{tabular}{lcdcc}
\multicolumn{1}{c}{Integral} &
\multicolumn{1}{c}{$n_F$} &
\multicolumn{1}{c}{\hspace*{4em}\parbox[t]{10em}{Value (Error) \\[-1ex] including $n_F$}} &
\multicolumn{1}{c}{\parbox[t]{10em}{Sampling per \\[-1ex]  iteration}} &
\multicolumn{1}{c}{\parbox[t]{4em}{No. of \\[-1ex] iterations}} \\[3ex]
\hline

$\Delta M_{A1}$ & 20 & -0.016~78~( 15) & $ 1\times 10^7,~1\times 10^8 $ & 50, 50\\
$\Delta M_{A2}$ & 20 & -0.004~71~(  8) & $ 1\times 10^7,~1\times 10^8 $ & 50, 50\\
$\Delta M_{A3}$ & 20 &  0.000~99~(  6) & $ 1\times 10^7,~1\times 10^8 $ & 50, 50\\
$\Delta M_{A4}$ & 10 & -0.003~93~(  1) & $ 1\times 10^7,~1\times 10^8 $ & 50, 50\\
$\Delta M_{A5}$ & 20 &  0.007~01~(  1) & $ 1\times 10^7,~1\times 10^8 $ & 50, 50\\
$\Delta M_{A6}$ & 20 & -0.023~43~( 12) & $ 1\times 10^7,~1\times 10^8 $ & 50, 50\\
$\Delta M_{A7}$ & 10 & -0.001~00~(  2) & $ 1\times 10^7,~1\times 10^8 $ & 50, 50\\
$\Delta M_{A8}$ & 10 &  0.001~97~(  1) & $ 1\times 10^7,~1\times 10^8 $ & 50, 50\\
$\Delta M_{B1}$ & 20 &  0.007~61~( 15) & $ 1\times 10^7,~1\times 10^8 $ & 50, 50\\
$\Delta M_{B2}$ & 20 &  0.000~37~(  8) & $ 1\times 10^7,~1\times 10^8 $ & 50, 50\\
$\Delta M_{B3}$ & 20 &  0.000~46~(  4) & $ 1\times 10^7,~1\times 10^8 $ & 50, 50\\
$\Delta M_{B4}$ & 10 &  0.003~05~(  1) & $ 1\times 10^7,~1\times 10^8 $ & 50, 50\\
$\Delta M_{B5}$ & 20 &  0.010~68~(  1) & $ 1\times 10^7,~1\times 10^8 $ & 50, 50\\
$\Delta M_{B6}$ & 20 &  0.015~17~( 11) & $ 1\times 10^7,~1\times 10^8 $ & 50, 50\\
$\Delta M_{B7}$ & 10 &  0.002~24~(  2) & $ 1\times 10^7,~1\times 10^8 $ & 50, 50\\
$\Delta M_{B8}$ & 10 & -0.013~72~(  1) & $ 1\times 10^7,~1\times 10^8 $ & 50, 50\\
$\Delta M_{C1}$ & 20 &  0.010~57~( 12) & $ 1\times 10^7,~1\times 10^8 $ & 50, 50\\
$\Delta M_{C2}$ & 20 &  0.004~88~(  5) & $ 1\times 10^7,~1\times 10^8 $ & 50, 50\\
$\Delta M_{C3}$ & 20 &  0.000~87~(  4) & $ 1\times 10^7,~1\times 10^8 $ & 50, 50\\
$\Delta M_{C4}$ & 10 &  0.000~84~(  1) & $ 1\times 10^7,~1\times 10^8 $ & 50, 50\\
$\Delta M_{C5}$ & 20 & -0.018~16~(  1) & $ 1\times 10^7,~1\times 10^8 $ & 50, 50\\
$\Delta M_{C6}$ & 20 &  0.009~57~( 10) & $ 1\times 10^7,~1\times 10^8 $ & 50, 50\\
$\Delta M_{C7}$ & 10 &  0.000~93~(  2) & $ 1\times 10^7,~1\times 10^8 $ & 50, 50\\
$\Delta M_{C8}$ & 10 &  0.011~40~(  1) & $ 1\times 10^7,~1\times 10^8 $ & 50, 50\\
\end{tabular}
\end{ruledtabular}
\end{table*}

\endgroup

\begingroup
\renewcommand{\baselinestretch}{1.0}

\begin{table*}
\caption{%
Contributions of diagrams of Set~III(c) to $a_\mu$
for $(l_1l_2) = (me)$.
The superscript $(me)$ is suppressed for simplicity.
The multiplicity $n_F$ is the number of vertex diagrams
represented by the integral and
is incorporated in the numerical value.
All integrals are evaluated initially with $10^8$ sampling points per iteration, iterated 50 times, followed by $10^9$ points, iterated several times.
  \label{table:setIII(c)_(me)}
}

\begin{ruledtabular}
\begin{tabular}{lcdcc}
\multicolumn{1}{c}{Integral} &
\multicolumn{1}{c}{$n_F$} &
\multicolumn{1}{c}{\hspace*{4em}\parbox[t]{10em}{Value (Error) \\[-1ex] including $n_F$}} &
\multicolumn{1}{c}{\parbox[t]{10em}{Sampling per \\[-1ex]  iteration}} &
\multicolumn{1}{c}{\parbox[t]{4em}{No. of \\[-1ex] iterations}} \\[3ex]
\hline

$\Delta M_{A1}$ & 20 & -18.722~( 37) & $ 1\times 10^8,~1\times 10^9 $ & 50,~376\\
$\Delta M_{A2}$ & 20 &  40.155~( 26) & $ 1\times 10^8,~1\times 10^9 $ & 50,~300\\
$\Delta M_{A3}$ & 20 &  -3.780~( 36) & $ 1\times 10^8,~1\times 10^9 $ & 50,~366\\
$\Delta M_{A4}$ & 10 & -18.309~( 11) & $ 1\times 10^8,~1\times 10^9 $ & 50,~ 30\\
$\Delta M_{A5}$ & 20 &   9.416~( 14) & $ 1\times 10^8,~1\times 10^9 $ & 50,~ 80\\
$\Delta M_{A6}$ & 20 & -37.911~( 30) & $ 1\times 10^8,~1\times 10^9 $ & 50,~301\\
$\Delta M_{A7}$ & 10 &  19.431~( 16) & $ 1\times 10^8,~1\times 10^9 $ & 50,~ 85\\
$\Delta M_{A8}$ & 10 &  10.371~(  7) & $ 1\times 10^8,~1\times 10^9 $ & 50,~ 70\\
$\Delta M_{B1}$ & 20 &  54.402~( 38) & $ 1\times 10^8,~1\times 10^9 $ & 50,~471\\
$\Delta M_{B2}$ & 20 & -73.374~( 29) & $ 1\times 10^8,~1\times 10^9 $ & 50,~300\\
$\Delta M_{B3}$ & 20 &  29.954~( 38) & $ 1\times 10^8,~1\times 10^9 $ & 50,~382\\
$\Delta M_{B4}$ & 10 &   2.578~( 13) & $ 1\times 10^8,~1\times 10^9 $ & 50,~ 30\\
$\Delta M_{B5}$ & 20 & -49.408~( 20) & $ 1\times 10^8,~1\times 10^9 $ & 50,~ 80\\
$\Delta M_{B6}$ & 20 &  -1.509~( 33) & $ 1\times 10^8,~1\times 10^9 $ & 50,~301\\
$\Delta M_{B7}$ & 10 &   9.521~( 20) & $ 1\times 10^8,~1\times 10^9 $ & 50,~ 85\\
$\Delta M_{B8}$ & 10 &  29.116~(  9) & $ 1\times 10^8,~1\times 10^9 $ & 50,~ 70\\
$\Delta M_{C1}$ & 20 & -31.212~( 37) & $ 1\times 10^8,~1\times 10^9 $ & 50,~497\\
$\Delta M_{C2}$ & 20 &  36.233~( 32) & $ 1\times 10^8,~1\times 10^9 $ & 50,~300\\
$\Delta M_{C3}$ & 20 & -25.285~( 37) & $ 1\times 10^8,~1\times 10^9 $ & 50,~409\\
$\Delta M_{C4}$ & 10 &  15.428~( 16) & $ 1\times 10^8,~1\times 10^9 $ & 50,~ 30\\
$\Delta M_{C5}$ & 20 &  28.857~( 22) & $ 1\times 10^8,~1\times 10^9 $ & 50,~ 80\\
$\Delta M_{C6}$ & 20 &  43.793~( 38) & $ 1\times 10^8,~1\times 10^9 $ & 50,~310\\
$\Delta M_{C7}$ & 10 & -27.637~( 17) & $ 1\times 10^8,~1\times 10^9 $ & 50,~ 85\\
$\Delta M_{C8}$ & 10 & -44.647~( 11) & $ 1\times 10^8,~1\times 10^9 $ & 50,~ 75\\

\end{tabular}
\end{ruledtabular}
\end{table*}

\endgroup

\begingroup
\renewcommand{\baselinestretch}{1.0}

\begin{table*}
\caption{%
Contributions of diagrams of Set~III(c) to $a_\mu$
for $(l_1l_2) = (mt)$.
The superscript $(mt)$ is suppressed for simplicity.
The multiplicity $n_F$ is the number of vertex diagrams
represented by the integral and
is incorporated in the numerical value.
All integrals are evaluated initially with $10^8$ sampling points per iteration, iterated 50 times, followed by $10^9$ points, iterated several times.
  \label{table:setIII(c)_(mt)}
}

\begin{ruledtabular}
\begin{tabular}{lcdcc}
\multicolumn{1}{c}{Integral} &
\multicolumn{1}{c}{$n_F$} &
\multicolumn{1}{c}{\hspace*{4em}\parbox[t]{10em}{Value (Error) \\[-1ex] including $n_F$}} &
\multicolumn{1}{c}{\parbox[t]{10em}{Sampling per \\[-1ex]  iteration}} &
\multicolumn{1}{c}{\parbox[t]{4em}{No. of \\[-1ex] iterations}} \\[3ex]
\hline

$\Delta M_{A1}$ & 20 & -0.423~43~(111) & $ 1\times 10^8,~1\times 10^9 $ & 50,~ 40\\
$\Delta M_{A2}$ & 20 & -0.001~66~( 64) & $ 1\times 10^8,~1\times 10^9 $ & 50,~ 30\\
$\Delta M_{A3}$ & 20 &  0.033~30~( 59) & $ 1\times 10^8,~1\times 10^9 $ & 50,~ 30\\
$\Delta M_{A4}$ & 10 & -0.102~91~( 13) & $ 1\times 10^8 $               & 50     \\
$\Delta M_{A5}$ & 20 &  0.327~57~( 14) & $ 1\times 10^8,~1\times 10^9 $ & 50,~ 15\\
$\Delta M_{A6}$ & 20 & -0.600~58~( 75) & $ 1\times 10^8,~1\times 10^9 $ & 50,~ 40\\
$\Delta M_{A7}$ & 10 &  0.011~26~( 30) & $ 1\times 10^8,~1\times 10^9 $ & 50,~ 15\\
$\Delta M_{A8}$ & 10 &  0.065~01~(  9) & $ 1\times 10^8,~1\times 10^9 $ & 50,~ 15\\
$\Delta M_{B1}$ & 20 &  0.246~29~(111) & $ 1\times 10^8,~1\times 10^9 $ & 50,~ 40\\
$\Delta M_{B2}$ & 20 & -0.104~08~( 64) & $ 1\times 10^8,~1\times 10^9 $ & 50,~ 30\\
$\Delta M_{B3}$ & 20 &  0.052~89~( 49) & $ 1\times 10^8,~1\times 10^9 $ & 50,~ 30\\
$\Delta M_{B4}$ & 10 &  0.076~46~( 11) & $ 1\times 10^8$                & 50     \\
$\Delta M_{B5}$ & 20 &  0.114~81~( 13) & $ 1\times 10^8,~1\times 10^9 $ & 50,~ 15\\
$\Delta M_{B6}$ & 20 &  0.397~59~( 67) & $ 1\times 10^8,~1\times 10^9 $ & 50,~ 40\\
$\Delta M_{B7}$ & 10 &  0.043~30~( 25) & $ 1\times 10^8,~1\times 10^9 $ & 50,~ 15\\
$\Delta M_{B8}$ & 10 & -0.298~83~(  9) & $ 1\times 10^8,~1\times 10^9 $ & 50,~ 15\\
$\Delta M_{C1}$ & 20 &  0.251~61~(108) & $ 1\times 10^8,~1\times 10^9 $ & 50,~ 40\\
$\Delta M_{C2}$ & 20 &  0.137~58~( 51) & $ 1\times 10^8,~1\times 10^9 $ & 50,~ 30\\
$\Delta M_{C3}$ & 20 & -0.052~89~( 51) & $ 1\times 10^8,~1\times 10^9 $ & 50,~ 30\\
$\Delta M_{C4}$ & 10 &  0.026~64~( 10) & $ 1\times 10^8 $               & 50     \\
$\Delta M_{C5}$ & 20 & -0.476~30~( 12) & $ 1\times 10^8,~1\times 10^9 $ & 50,~ 15\\
$\Delta M_{C6}$ & 20 &  0.268~81~( 65) & $ 1\times 10^8,~1\times 10^9 $ & 50,~ 40\\
$\Delta M_{C7}$ & 10 & -0.038~53~( 25) & $ 1\times 10^8,~1\times 10^9 $ & 50,~ 15\\
$\Delta M_{C8}$ & 10 &  0.213~27~( 14) & $ 1\times 10^8 $               & 50     \\
\end{tabular}
\end{ruledtabular}
\end{table*}

\endgroup

\section{Discussion}
\label{sec:discussion}

All programs of diagrams of the Set III(c) were written 
in two independent ways, in order to detect possible programming error.
No such error was found.

The value of $A_2^{(10)} [\text{Set III(c)}^{(me)}]$ given in (\ref{set3c_me})
is not much larger than that of
$A_1^{(10)} [\text{Set III(c)}^{(ee)}]$ given in (\ref{set3c_ee}).
This is somewhat unexpected since,
as is seen from Table \ref{table:setIII(c)_(me)}, individual integrals contributing to
$A_2^{(10)} [\text{Set III(c)}^{(me)}]$ are an order of magnitude larger than
those given in Table \ref{table:setIII(c)_(ee)}.
Presumably, the modest value of (\ref{set3c_me})
is a consequence of strong cancellation among contributing integrals.

\begin{acknowledgments}
We thank Mr.~N.~Watanabe for his contribution 
in the early stage of  this work.
This work is supported in part by the JSPS Grant-in-Aid for Scientific 
Research (C)19540322, (C)20540261, and (C)23540331.
The part of material presented by T.~K. is based on work supported
by the U. S. National Science Foundation under the Grant NSF-PHY-0757868,
and the International Exchange Support Grants (FY2010) of RIKEN.
T.~K. thanks RIKEN for the hospitality extended to him while
a part of this work was carried out.
Numerical calculations  are conducted 
on the RIKEN Supercombined Cluster System (RSCC),
the RIKEN Integrated Cluster of Clusters (RICC) supercomputing systems,
and the  $\varphi$ computer of Kobayashi-Maskawa Institute.
\end{acknowledgments}


\appendix

\section{Renormalization of diagrams of Set III(c)}
\label{sec:renorm}

Diagrams of Set III(c), shown in Figure~\ref{fig:set3c},
contain an {\em l-l} subdiagram internally. 
Thus we find it convenient to pursue a renormalization
scheme somewhat different from all other Sets
contributing to the tenth-order $g\!-\!2$.
As is indicated by figures $A1$, $B1$, $C1$ of Figure~ \ref{fig:set3c},
we denote open fermion lines as 1, 2, 3, 4, 5,
closed fermion lines as 6, 7, 8, 9, and  photon lines  as
a, b, c, d, e.
We will identify a subdiagram containing open lepton lines
in terms of their line numbers.
For instance, the second-order vertex subdiagram \{1,2;e\} 
and sixth-order vertex subdiagram \{4,5;6,7,8,9;b,c,d\} 
of $A1$ will be 
denoted by the superscript (1,2) and (4,5), respectively.
An exception is the {\em l-l} subdiagram, which will be denoted as (6,7,8,9). 
Of course this is just for the sake of keeping track of where
a particular subdiagram is located.
The superscript will be removed when it is no longer needed.

\subsection{\textit{A1, B1, C1}}
\label{sec:A1B1C1}

Let us begin with the $g\!-\!2$ amplitude $M_{A1}$.
Noting that, out of 15 forests
of the diagram $A1$  mentioned in Sec.~\ref{renormalization},
8 are hidden in our convention leading to Eq. (\ref{FminusG}),
the renormalized amplitude $a_{A1}$ can be written as
\begin{eqnarray}
a_{A1} &=& M_{A1} 
        - L_2^{(1,2)} M_{8LLJ}^{(3,4,5)}           
        - L_{6LL(5)}^{(4,5)} M_{4a}^{(1,2,3)}      
        - L_{8LLF(7)}^{(1,2,3,4)} M_2^{(5)}        
        - L_{8LLJ(1)}^{(2,3,4,5)} M_2^{(1)}        
\nonumber \\
       &+& L_2^{(1,2)} L_{6LL(5)}^{(4,5)} M_2^{(3)}  
        +  L_2^{(1,2)} L_{6LL(5)}^{(3,4)} M_2^{(5)}  
        +  L_2^{(2,3)} L_{6LL(5)}^{(4,5)} M_2^{(1)}. 
\label{a_A1}
\end{eqnarray}
As was discussed in Sec.~\ref{sec:formulation},
all terms of (\ref{a_A1}) containing an {\em l-l} subdiagram 
are to be understood as shorthands for the 
regularized quantity defined by Eq. (\ref{FminusG}).
In other words, the UV divergence arising from 
the {\em l-l} subdiagram has been removed
by the procedure described in Sec.~\ref{sec:formulation}  so that it can be
treated as a UV-finite quantity.
$M_{8LLJ}$ is the proper magnetic moment amplitude of the eighth-order
diagram $8LLJ$ of Figure~\ref{fig:LLjkl}. 
See \cite{Kinoshita:1981ww,Kinoshita:2002ns} for its precise definition.
$L_2$ is the vertex renormalization constant of the second order.
$L_{6LL(5)}$ is the renormalization constant associated with
the sixth-order vertex diagram $6LL(5)$ shown in Figure~\ref{fig:6LL}.
$L_{8LLF(7)}$ and $L_{8LLJ(1)}$ are the eighth-order vertex 
renormalization constants associated with 
the self-energy-like diagrams $8LLF$ of Figure~\ref{fig:G4c}
and $8LLJ$  of Figure~\ref{fig:LLjkl}, respectively.

In the amplitude $M_{A1}$, the ${\mathit K}$-operation is applied only on 
the {\it explicit} second-order vertex subdiagram (1,2). 
For other terms the full bodies of the vertex renormalization
constants of the sixth- and eighth-orders are used and subtracted.
These vertex renormalization constants are extracted  from a 
vertex diagram $\Gamma_\nu(p,q)$, where $(p-q/2)^2 =(p+q/2)^2=m^2$, 
using the projection operator
\begin{equation}
L = \frac{1}{4} \mathrm{Tr} [ (\pslash + m ) p^\nu \Gamma_\nu ] |_{q = 0} ~.
\end{equation}
The result is combined with the lower-order magnetic moment amplitude
using, for instance, the factorization procedure described in Sec.~III~D of
Ref.~\cite{Aoyama:2005kf} backwards so that the combined
formula is described by the same set of Feynman parameters as those of
the unrenormalized magnetic moment $M_{A1}$. 
Then the UV-finite quantity $\Delta M_{A1}$ can be written as
\begin{eqnarray}
\Delta M_{A1} &=& M_{A1}
        - L_2^{UV(1,2)} M_{8LLJ}^{(3,4,5)}           
        - L_{6LL(5)}^{(4,5)} M_{4a}^{(1,2,3)}        
        - L_{8LLF(7)}^{(1,2,3,4)} M_2^{(5)}          
        - L_{8LLJ(1)}^{(2,3,4,5)} M_2^{(1)}          
\nonumber \\
       &+& L_2^{UV(1,2)} L_{6LL(5)}^{(3,4)} M_2^{(5)}  
        +  L_2^{UV(1,2)} L_{6LL(5)}^{(4,5)} M_2^{(3)}  
        +  L_2^{(2,3)} L_{6LL(5)}^{(4,5)} M_2^{(1)}.   
\label{delA_A1}
\end{eqnarray}
where $L_2^{UV(1,2)}$ is the UV-divergent part of $L_2^{(1,2)}$
defined by the $K$-operation.
Note that
        $L_{6LL(5)}^{(3,4)} $,
        $L_{6LL(5)}^{(4,5)} $,
        $L_{8LLF(7)}^{(1,2,3,4)} $,
        $L_{8LLJ(1)}^{(2,3,4,5)} $, 
        $L_2^{(2,3)}$, and $L_{6LL(5)}^{(4,5)}$
are not decomposed into UV-divergent and UV-finite parts.
Note, in particular, that $L_2^{(2,3)}$ has an IR-divergent part 
besides a UV-divergent part. 
This is the reason why we normally  avoid use of
the whole $L_2$ as a subtraction term and
use the $L_2^{UV}$ defined by  the ${\mathit K}$-operation instead.
For the diagram $A1$, however, the IR divergences in $L_2^{(2,3)}$  
and $L_{8LLJ(1)}$ cancel each other.
Thus the UV-divergence-free amplitude  $\Delta M_{A1}$ is 
also IR-divergence-free.
The numerical integration code for the Set III(c) is constructed
taking this observation into account.

Substituting (\ref{delA_A1}) in (\ref{a_A1}) we obtain
\begin{eqnarray}
a_{A1} &=& \Delta M_{A1} - L^{\rm R}_2 M_{8LLJ}^{(3,4,5)} 
                         + L^{\rm R}_2 L_{6LL(5)}^{(3,4)} M_2 
                         + L^{\rm R}_2 L_{6LL(5)}^{(4,5)} M_2 
\nonumber \\
       &=& \Delta M_{A1} - L^{\rm R}_2 \Delta M_{8LLJ},
\label{finala_A1}
\end{eqnarray}
where $L^{\rm R}_2 \equiv L_2 - L_2^{UV}$ is UV-finite but IR-divergent
and $\Delta M_{8LLJ} = M_{8LLJ} - 2 L_{6LL(5)}M_2 $ 
is the finite $g\!-\!2$ contribution from the eighth-order diagram $8LLJ$ 
\cite{Kinoshita:1981ww,Kinoshita:2002ns}.

Similar consideration for the diagrams $B1$ and $C1$ yields
\begin{eqnarray}
a_{B1} &=& \Delta M_{B1} - L^{\rm R}_2 M_{8LLL}^{(3,4,5)} + 2 L^{\rm R}_2 L_{6LL(3)} M_2 
\nonumber \\
       &=& \Delta M_{B1} - L^{\rm R}_2 \Delta M_{8LLL},
\label{finala_B1}
\end{eqnarray}
and
\begin{eqnarray}
a_{C1} &=& \Delta M_{C1} - L^{\rm R}_2 M_{8LLK}^{(3,4,5)} + 2 L^{\rm R}_2 L_{6LL(3)}^{(3,4)} M_2
\nonumber \\
       &=& \Delta M_{C1} - L^{\rm R}_2 \Delta M_{8LLK}.
\label{finala_C1}
\end{eqnarray}
%
From (\ref{finala_A1}), (\ref{finala_B1}), and (\ref{finala_C1}) we obtain
\begin{equation}
\sum_{\alpha = A}^C a_{\alpha 1} =
\sum_{\alpha = A}^C \Delta M_{\alpha 1} - L^{\rm R}_2 \Delta M_{8JKL},
\label{A1B1C1}
\end{equation}
where $\Delta M_{8JKL} \equiv \Delta M_{8LLJ}+\Delta M_{8LLK}+\Delta M_{8LLL}$ \cite{Kinoshita:1981ww}.
Note that the sum $L_{6LL} \equiv L_{6LL(3)} +L_{6LL(4)} +L_{6LL(5)}=0 $ 
because of the gauge invariance.

\subsection{\textit{A2, B2, C2}}
\label{sec:A2B2C2}

The diagram $A2$ has UV-divergent subdiagrams $(1,2,3,4)$, $(2,3,4,5)$
besides the {\em l-l} subdiagram $(6,7,8,9)$.
Thus the renormalized amplitude $a_{A2}$ can be written as
\begin{equation}
a_{A2} = M_{A2} -  M_2^{(1)} L_{8LLJ(2)}^{(2,3,4,5)} - L_{8LLG(7)}^{(1,2,3,4)} M_2^{(5)}.
\label{a_A2}
\end{equation}
Diagrams $8LLJ$ and $8LLG$ are shown in Figure~\ref{fig:G4c}.
Since $A2$ has no UV divergence due to the second-order subdiagram,
we define $\Delta M_{A2}$ by
\begin{equation}
\Delta M_{A2} = M_{A2} - M_2^{(1)} L_{8LLJ(2)}^{(2,3,4,5)} - L_{8LLG(7)}^{(1,2,3,4)} M_2^{(5)}.
\label{delA_A2}
\end{equation}
Substituting (\ref{delA_A2}) in (\ref{a_A2}) we obtain
\begin{equation}
a_{A2} = \Delta M_{A2} .
\label{finala_A2}
\end{equation}
Similar equations hold for $a_{B2}$ and $a_{C2}$.  Thus we have
\begin{equation}
\sum_{\alpha = A}^C a_{\alpha 2} =
\sum_{\alpha = A}^C \Delta M_{\alpha 2}.
\label{A2B2C2}
\end{equation}

\subsection{\textit{A3, B3, C3}}
\label{sec:A3B3C3}

The diagram $A3$ has five forests after the {\em l-l} subdiagrams are treated
following the consideration of Sec.~\ref{renormalization}.
Thus the renormalized amplitude $a_{A3}$ can be written as
\begin{eqnarray}
a_{A3} &=& M_{A3} -  M_2^{(1)} L_{8LLJ(3)}^{(2,3,4,5)}  
                  -  M_2^{(5)} L_{8LLH(7)}^{(1,2,3,4)}  
                -  M_{4a}^{(1,4,5)} L_{6LL(5)}^{(2,3)}  
\nonumber \\
                &+&  M_{2}^{(1)}L_2^{(4,5)} L_{6LL(5)}^{(2,3)}  
                +  M_{2}^{(5)}L_2^{(1,4)} L_{6LL(5)}^{(2,3)}  
                .
\label{a_A3}
\end{eqnarray}
The second-order vertex renormalization constants $L_2^{(1,4)}$
and $L_2^{(4,5)}$ appear in (\ref{a_A3}) as reduced diagrams,
which we called {\it implicit}, and used the full
renormalization constant $L_2$ for them. 
Thus we define the finite amplitude by
\begin{eqnarray}
\Delta M_{A3} &=&    M_{A3} -  M_2^{(1)} L_{8LLJ(3)}^{(2,3,4,5)}  
                  -  M_2^{(5)} L_{8LLH(7)}^{(1,2,3,4)}  
                -  M_{4a}^{(1,4,5)} L_{6LL(5)}^{(2,3)}  
\nonumber \\
                &+&  M_{2}^{(1)}L_2^{(4,5)} L_{6LL(5)}^{(2,3)}  
                +  M_{2}^{(5)}L_2^{(1,4)} L_{6LL(5)}^{(2,3)}. 
\label{delA_A3}
\end{eqnarray}
In other words, we have
\begin{equation}
a_{A3} = \Delta M_{A3}. 
\label{finala_A3}
\end{equation}
Similar relation holds for $a_{B3}$ and $a_{C3}$.  Thus we have
\begin{equation}
\sum_{\alpha = A}^C a_{\alpha 3} =
\sum_{\alpha = A}^C \Delta M_{\alpha 3}.
\label{A3B3C3}
\end{equation}
%

\subsection{\textit{A4, B4, C4}}
\label{sec:A4B4C4}

The diagram $A4$ has one self-energy subdiagram (2,3,4)
and two vertex subdiagrams (2,3) and (3,4) 
as well as the {\em l-l} subdiagram (6,7,8,9).
Thus the renormalized amplitude $a_{A4}$ is given by
\begin{eqnarray}
a_{A4} &=& M_{A4}- M_{4b}^{(1,2,5)} L_{6LL(5)}^{(3,4)}  
                 - M_{4b}^{(1,4,5)} L_{6LL(5)}^{(2,3)}  
                 - M_2^{(1,5)} B_{8LLJ}^{(2,3,4)}
                 - M_{2^*}^{(1,5)} \delta m_{8LLJ}^{(2,3,4)}
\nonumber \\
               &+& M_2^{(1,5)} B_2^{(2)} L_{6LL(5)}^{(3,4)}
                +  M_{2^*}^{(1,5)} \delta m_2^{(2)} L_{6LL(5)}^{(3,4)}
\nonumber \\
               &+& M_2^{(1,5)} B_2^{(4)} L_{6LL(5)}^{(2,3)}
                +  M_{2^*}^{(1,5)} \delta m_2^{(4)} L_{6LL(5)}^{(2,3)}.
\label{a_A4}
\end{eqnarray}
We define the UV-finite amplitude $\Delta^{\prime} M_{A4}$ by
\begin{eqnarray}
\Delta^{\prime} M_{A4} &=& M_{A4}
               - M_{4b}^{(1,2,5)} L_{6LL(5)}^{(3,4)}  
               - M_{4b}^{(1,4,5)} L_{6LL(5)}^{(2,3)}  
               - M_2^{(1,5)} B_{8LLJ}^{(2,3,4)}
               - M_{2^*}^{(1,5)} \delta m_{8LLJ}^{(2,3,4)}
\nonumber \\
              &+& M_2^{(1,5)} B_{2}^{(2)} L_{6LL(5)}^{(3,4)}
               + M_{2^*}^{(1,5)} \delta m_2^{(2)} L_{6LL(5)}^{(3,4)}
\nonumber \\
              &+& M_2^{(1,5)} B_{2}^{(4)} L_{6LL(5)}^{(2,3)}
               + M_{2^*}^{(1,5)} \delta m_2^{(4)} L_{6LL(5)}^{(2,3)},
\label{delA_A4}
\end{eqnarray}
where $M_{2^*}$ is derived from $M_2$ by inserting a two-point vertex
in the lepton line.
Note that the renormalization constants $B_2$ and $\delta m_2$ arising from
the self-energy subdiagrams \{2;a\} and \{4;d\}
are subtracted as a whole
without breaking them up into UV-divergent and UV-finite parts. 
This is consistent only if we use the full body of the renormalization 
constant $B_{8LLJ}$. Otherwise, IR-singular part of $B_{8LLJ}$ and 
two $B_2$'s do not cancel out each other.
Substituting (\ref{delA_A4}) in (\ref{a_A4}), we obtain
\begin{equation}
a_{A4} = \Delta^{'} M_{A4}. 
\label{finala_A4}
\end{equation}

The resulting $a_{A4}$ is UV-finite but IR-divergent.
Separating the IR divergence of $\Delta^{'} M_{A4}$ 
from the subdiagram \{1,5;e\} by the ${\mathit I}$-operation, 
we can write
\begin{equation}
a_{A4} = \Delta M_{A4} + L^{\rm R}_2 (M_{8LLJ}^{(2,3,4)}-2 L_{6LL(5)} M_2) 
       = \Delta M_{A4} + L^{\rm R}_2 \Delta M_{8LLJ}^{(2,3,4)} .
\label{da_A4}
\end{equation}
Similarly we have
\begin{equation}
a_{B4} = \Delta M_{B4} + L^{\rm R}_2 (M_{8LLL}^{(2,3,4)}-2 L_{6LL(3)} M_2) 
       = \Delta M_{B4} + L^{\rm R}_2 \Delta M_{8LLL}^{(2,3,4)},
\label{da_B4}
\end{equation}
and
\begin{equation}
a_{C4} = \Delta M_{B4} + L^{\rm R}_2 (M_{8LLK}^{(2,3,4)}-2 L_{6LL(3)} M_2) 
       = \Delta M_{B4} + L^{\rm R}_2 \Delta M_{8LLK}^{(2,3,4)} .
\label{da_C4}
\end{equation}
Adding up these three results we obtain
\begin{equation}
\sum_{\alpha = A}^C a_{\alpha 4} =
\sum_{\alpha = A}^C \Delta M_{\alpha 4} + L^{\rm R}_2 \Delta M_{8JKL},
\label{A4B4C4}
\end{equation}
noting that gauge invariance guarantees the vanishing of
the sum $L_{6LL} \equiv L_{6LL(3)} +L_{6LL(4)} +L_{6LL(5)}=0 $.

We also developed an alternative method for separating UV-divergence
from $M_{A4}$, in which a UV-finite amplitude is defined by
\begin{eqnarray}
\Delta^{\prime \prime} M_{A4} &=& M_{A4}
               - M_{4b}^{(1,2,5)} L_{6LL(5)}^{(3,4)}  
               - M_{4b}^{(1,4,5)} L_{6LL(5)}^{(2,3)}  
               - M_2^{(1,5)} B_{8LLJ}(E)^{(2,3,4)}
               - M_{2^*}^{(1,5)} \delta m_{8LLJ}^{(2,3,4)}
\nonumber \\
              &+& M_2^{(1,5)} B_{2}^{(2)}(E)  L_{6LL(5)}^{(3,4)}
               + M_{2^*}^{(1,5)} \delta m_2^{(2)} L_{6LL(5)}^{(3,4)}
\nonumber \\
              &+& M_2^{(1,5)} B_{2}^{(4)}(E) L_{6LL(5)}^{(2,3)}
               + M_{2^*}^{(1,5)} \delta m_2^{(4)} L_{6LL(5)}^{(2,3)} ,
\label{delA_A4_alternative}
\end{eqnarray}
where
\begin{eqnarray}
              B_{8LLJ} &=& B_{8LLJ}(E) + B_{8LLJ}(N) ,
              \nonumber \\
              B_2&=& B_2(E) + B_2(N).
\end{eqnarray}
The $B(E)$ term of the wave-function renormalization constant
comes from the derivative of the numerator of the self-energy diagram
$\Sigma(p)$  with respect to the fermion momentum $p$, while
the $B(N)$ term is the derivative of the denominator function $V$ 
defined in (\ref{eq:defV}).
For the second-order case, we find $B_2(E) = B_2^{\rm UV}$ 
and $B_2(N) = B_2^{\rm R}$.
The relationship to the fully subtracted $\Delta^{\prime} M_{A4}$ is
thus clear and we find
\begin{eqnarray}
\Delta^{\prime } M_{A4} &=    & \Delta^{\prime \prime} M_{A4} 
                         - M_2^{(1,5)} \Delta B_{8LLJ}^{(2,3,4)} ,
\nonumber \\
   \Delta B_{8LLJ} &\equiv & B_{8LLJ}(N) - 2 B_2^{\rm R} L_{6LL(5)} ~.
\end{eqnarray}
The IR subtraction term used for $\Delta^{\prime \prime} M_{A4}$ is
the same one for $\Delta^{\prime } M_{A4}$.
As a check we evaluated both integrals numerically.
The results are in good agreement within the uncertainty of VEGAS 
integration.


\subsection{\textit{A5, B5, C5}}
\label{sec:A5B5C5}

The diagram $A5$ has a self-energy subdiagram (2)
and two vertex subdiagrams (1,2,3,4) and (4,5)
besides the {\em l-l} subdiagram (6,7,8,9).
The subdiagram (2) is the second-order self-energy diagram
which contributes to the renormalization constants 
$B_2$ and $\delta m_2$. Taking this into account we can write
the renormalized amplitude $a_{A5}$ as
\begin{eqnarray}
a_{A5} &=& M_{A5} 
       - L_{6LL(5)}^{(4,5)} M_{4b}^{(1,2,3)}
       - L_{8LLI(7)}^{(1,2,3,4)} M_2^{(5)}
       - B_2^{(2)} M_{8LLJ}^{(1,3,4,5)}
       - \delta m_2^{(2)} M_{8LLJ^*}^{(1,3,4,5)}
\nonumber \\
      &+& B_2^{(2)} L_{6LL(5)}^{(1,3,4)} M_2^{(5)}
       + \delta m_2^{(2)} L_{6LL^*(5)}^{(1,3,4)} M_2^{(5)}
\nonumber \\
      &+& B_2^{(2)} L_{6LL(5)}^{(4,5)} M_2^{(1,3)}
       + \delta m_2^{(2)} L_{6LL(5)}^{(4,5)} M_{2^*}^{(1,3)} .
\label{a_A5}
\end{eqnarray}
Applying the ${\mathit K}_2$-operation to the self-energy  subdiagram (2),
we obtain
\begin{eqnarray}
\Delta M_{A5} &=& M_{A5} 
       - L_{6LL(5)}^{(4,5)} M_{4b}^{(1,2,3)}
       - L_{8LLI(7)}^{(1,2,3,4)} M_2^{(5)}
       - B_2^{(2)UV} M_{8LLJ}^{(1,3,4,5)}
       - \delta m_2^{(2)} M_{8LLJ^*}^{(1,3,4,5)}
\nonumber \\
       &+& B_2^{(2)UV} L_{6LL(5)}^{(1,3,4)} M_2^{(5)}
       + \delta m_2^{(2)} L_{6LL^*(5)}^{(1,3,4)} M_2^{(5)}
\nonumber \\
       &+& B_2^{(2)UV} L_{6LL(5)}^{(4,5)} M_2^{(1,3)}
       + \delta m_2^{(2)} L_{6LL(5)}^{(4,5)} M_{2^*}^{(1,3)} .
\label{delA_A5}
\end{eqnarray}
Note that the ${\mathit K}_2$-operation yields the whole mass-renormalization
constant $\delta m_2$.
Substituting (\ref{delA_A5}) in (\ref{a_A5}), we obtain
\begin{eqnarray}
a_{A5}&=& \Delta M_{A5} 
       - B^{\rm R}_2 M_{8LLJ}^{(1,3,4,5)} 
       + B^{\rm R}_2 L_{6LL(5)}^{(1,3,4)} M_2^{(5)}
       + B^{\rm R}_2 L_{6LL(5)}^{(4,5)} M_2^{(1,3)}
\nonumber \\
      &=& \Delta M_{A5} - B^{\rm R}_2 \Delta M_{8LLJ}
\label{finala_A5}
\end{eqnarray}
where $B^{\rm R}_2 = B_2 - B_2^{UV}$.
Similar consideration for the diagrams B5 and C5 yields
\begin{eqnarray}
a_{B5}&=& \Delta M_{B5} 
       - B^{\rm R}_2 M_{8LLL}^{(1,3,4,5)} 
       + B^{\rm R}_2 L_{6LL(3)}^{(1,3,4)} M_2^{(5)}
       + B^{\rm R}_2 L_{6LL(3)}^{(4,5)} M_2^{(1,3)}
\nonumber \\
      &=& \Delta M_{B5} - B^{\rm R}_2 \Delta M_{8LLL}
\label{finala_B5}
\\
a_{C5}&=& \Delta M_{C5}
       - B^{\rm R}_2 M_{8LLK}^{(1,3,4,5)} 
       + B^{\rm R}_2 L_{6LL(3)}^{(1,3,4)} M_2^{(5)}
       + B^{\rm R}_2 L_{6LL(3)}^{(4,5)} M_2^{(1,3)}
\nonumber \\
      &=&\Delta M_{C5} - B^{\rm R}_2 \Delta M_{8LLK}
        .
\label{finala_C5}
\end{eqnarray}
Adding up these results, we obtain
\begin{equation}
\sum_{\alpha = A}^C a_{\alpha 5} =
\sum_{\alpha = A}^C \Delta M_{\alpha 5} - B^{\rm R}_2 \Delta M_{8JKL}.
\label{A5B5C5}
\end{equation}
%

\subsection{\textit{A6, B6, C6}}
\label{sec:A6B6C6}

The diagram $A6$ has UV-divergent subdiagrams
$(2,3), (1,2,3,4), (2,3,4,5)$, besides $(6,7,8,9)$,
and the corresponding forest structure.
Thus the renormalized amplitude $a_{A6}$ can be written as
\begin{eqnarray}
a_{A6} &=& M_{A6} 
        - L_2^{(2,3)} M_{8LLJ}^{(1,4,5)}           
        - L_{8LLE(7)}^{(1,2,3,4)} M_2^{(5)}        
        - L_{8LLF(7)}^{(2,3,4,5)} M_2^{(1)}        
\nonumber \\
       &+& L_2^{(2,3)} L_{6LL(5)}^{(4,5)} M_2^{(1)}  
        +  L_2^{(2,3)} L_{6LL(5)}^{(1,4)} M_2^{(5)}. 
\label{a_A6}
\end{eqnarray}

Applying the ${\mathit K}_{23}$-operation on $M_{A6}$, we can define 
the UV-finite quantity $\Delta M_{A6}$ as
\begin{eqnarray}
\Delta M_{A6} &=& M_{A6}
        - L_2^{UV(2,3)} M_{8LLJ}^{(1,4,5)}           
        - L_{8LLE(7)}^{(1,2,3,4)} M_2^{(5)}          
        - L_{8LLF(7)}^{(2,3,4,5)} M_2^{(1)}          
\nonumber \\
       &+&  L_2^{UV(2,3)} L_{6LL(5)}^{(4,5)} M_2^{(1)}  
        +  L_2^{UV(2,3)} L_{6LL(5)}^{(1,4)} M_2^{(5)}.   
\label{delA_A6}
\end{eqnarray}

Substituting (\ref{delA_A6}) in (\ref{a_A6}), we obtain
\begin{eqnarray}
a_{A6} &=& \Delta M_{A6} - L^{\rm R}_2 M_{8LLJ}^{(1,4,5)} +
 L^{\rm R}_2 L_{6LL(5)}^{(4,5)} M_2 + L^{\rm R}_2 L_{6LL(5)}^{(1,4)} M_2
\nonumber \\
       &=& \Delta M_{A6} - L^{\rm R}_2 \Delta M_{8LLJ}.
\label{finala_A6}
\end{eqnarray}

Similar consideration for the diagrams $B6$ and $C6$ yields
\begin{eqnarray}
a_{B6} &=& \Delta M_{B6} - L^{\rm R}_2 M_{8LLL}^{(1,4,5)} 
+  L^{\rm R}_2 L_{6LL(3)}^{(4,5)} M_2 
+  L^{\rm R}_2 L_{6LL(3)}^{(1,4)} M_2 
\nonumber \\
      & =& \Delta M_{B6} - L^{\rm R}_2 \Delta M_{8LLL},
\label{finala_B6}
\end{eqnarray}
and
\begin{eqnarray}
a_{C6} &=& \Delta M_{C6} - L^{\rm R}_2 M_{8LLK}^{(1,4,5)} 
+  L^{\rm R}_2 L_{6LL(3)}^{(4,5)} M_2
+  L^{\rm R}_2 L_{6LL(3)}^{(1,4)} M_2
\nonumber \\ 
       &=& \Delta M_{C6} - L^{\rm R}_2 \Delta M_{8LLK}.
\label{finala_C6}
\end{eqnarray}
%
From (\ref{finala_A6}), (\ref{finala_B6}), and (\ref{finala_C6}), we obtain
\begin{equation}
\sum_{\alpha = A}^C a_{\alpha 6} =
\sum_{\alpha = A}^C \Delta M_{\alpha 6} - L^{\rm R}_2 \Delta M_{8JKL}.
\label{A6B6C6}
\end{equation}

\subsection{\textit{A7, B7, C7}}
\label{sec:A7B7C7}

The diagram $A7$ has two vertex subdiagrams (1,2,3,4) and (2,3,4,5),
besides the {\em l-l} subdiagram (6,7,8,9).
Thus the renormalized amplitude $a_{A7}$ can be written as
\begin{eqnarray}
a_{A7} &=& M_{A7} 
        - L_{8LLG(7)}^{(1,2,3,4)} M_2^{(5)}   
        - L_{8LLG(7)}^{(2,3,4,5)} M_2^{(1)}.    
\label{a_A7}
\end{eqnarray}
We define the UV-finite quantity $\Delta M_{A7}$ by
\begin{eqnarray}
\Delta M_{A7} &=& M_{A7} 
        - L_{8LLG(7)}^{(1,2,3,4)} M_2^{(5)} 
        - L_{8LLG(7)}^{(2,3,4,5)} M_2^{(1)}, 
\label{delA_A7}
\end{eqnarray}
where $L_{8LLG(7)}$ is not decomposed into UV-divergent and UV-finite parts.
Thus we have
\begin{equation}
a_{A7} = \Delta M_{A7}. 
\label{finala_A7}
\end{equation}
Similar relation holds for $a_{B7}$ and $a_{C7}$.  Thus we have
\begin{equation}
\sum_{\alpha = A}^C a_{\alpha 7} =
\sum_{\alpha = A}^C \Delta M_{\alpha 7}.
\label{A7B7C7}
\end{equation}
%


\subsection{\textit{A8, B8, C8}}
\label{sec:A8B8C8}

The diagram $A8$ has a self-energy subdiagram (3)
and two vertex subdiagrams (1,2,3,4) and (2,3,4,5),
besides the {\em l-l} subdiagram (6,7,8,9).
Thus, its renormalization structure is similar to that of the diagram $A5$:
\begin{eqnarray}
a_{A8} &=& M_{A8} 
       - L_{8LLI(7)}^{(1,2,3,4)} M_2^{(5)}
       - L_{8LLI(7)}^{(2,3,4,5)} M_2^{(1)}
       - B_2^{(3)} M_{8LLJ}^{(1,2,4,5)}
       - \delta m_2^{(3)} M_{8LLJ^*}^{(1,2,4,5)}
\nonumber \\
      &+&B_2^{(3)} L_{6LL(5)}^{(1,2,4)} M_2^{(5)}
       +\delta m_2^{(3)} L_{6LL^*(5)}^{(1,2,4)} M_2^{(5)}
\nonumber \\
      &+&B_2^{(3)} L_{6LL(5)}^{(2,4,5)} M_2^{(1)}
       +\delta m_2^{(3)} L_{6LL^*(5)}^{(2,4,5)} M_2^{(1)}
       .
\label{a_A8}
\end{eqnarray}
Applying the ${\mathit K}_3$-operation to the self-energy  subdiagram (3),
we obtain
\begin{eqnarray}
\Delta M_{A8} &=& M_{A8} 
       - L_{8LLI(7)}^{(1,2,3,4)} M_2^{(5)}
       - L_{8LLI(7)}^{(2,3,4,5)} M_2^{(1)}
       - B_2^{(3)UV} M_{8LLJ}^{(1,2,4,5)}
       - \delta m_2^{(3)} M_{8LLJ^*}^{(1,2,4,5)}
\nonumber \\
      &+&B_2^{(3)UV} L_{6LL(5)}^{(1,2,4)} M_2^{(5)}
       +\delta m_2^{(3)} L_{6LL^*(5)}^{(1,2,4)} M_2^{(5)}
\nonumber \\
      &+&B_2^{(3)UV} L_{6LL(5)}^{(2,4,5)} M_2^{(1)}
       +\delta m_2^{(3)} L_{6LL^*(5)}^{(2,4,5)} M_2^{(1)}
        .
\label{delA_A8}
\end{eqnarray}
Substituting (\ref{delA_A8}) in (\ref{a_A8}), we obtain
\begin{eqnarray}
a_{A8}&=& 
     \Delta M_{A8} - B^{\rm R}_2 \Delta M_{8LLJ}^{(1,3,4,5)} 
                   + B^{\rm R}_2 L_{6LL(5)}^{(1,2,4)} \Delta M_{2}^{(5)} 
                   + B^{\rm R}_2 L_{6LL(5)}^{(2,4,5)} \Delta M_{2}^{(1)} 
\nonumber \\
     &=& \Delta M_{A8} - B^{\rm R}_2 \Delta M_{8LLJ}^{(1,3,4,5)} 
     .
\label{finala_A8}
\end{eqnarray}

Applying the same consideration to the diagrams
$B8$ and $C8$, and adding them to (\ref{finala_A8}), we obtain
\begin{equation}
\sum_{\alpha = A}^C a_{\alpha 8} =
\sum_{\alpha = A}^C \Delta M_{\alpha 8} - B^{\rm R}_2 \Delta M_{8JKL}.
\label{A8B8C8}
\end{equation}
%

\subsection{Sum}
\label{sec:Sum}

Taking into account that
integrals for diagrams such as $A1$ actually represent $2 \times 2 \times 5$ vertex diagrams,
the sum of all diagrams of Set III(c)
can be written as
\begin{equation}
A_1^{(10)} [\text{Set~III(c)}^{(l_1l_2)}] 
 = \sum_{\alpha = A}^C \sum_{i=1}^8 \eta_i \Delta M_{\alpha i}^{(l_1l_2)}
        -3 \Delta L\!B_2 \Delta M_{8JKL}^{(l_1l_2)} ,
\end{equation}
where $l_1$ refers to the open lepton line and $l_2$ refers to 
the closed lepton line.
$\Delta LB_2 \equiv L^{\rm R}_2 + B^{\rm R}_2$,
and $\eta_i = 2 $ for $i=4, 7, 8$, $\eta_i = 4 $ for $i=1, 2, 3, 5, 6$.


\bibliographystyle{apsrev}
\bibliography{b}

\end{document}